\documentclass[twoside,twocolumn,9pt]{article}
\usepackage{extsizes}
\usepackage[super,sort&compress,comma]{natbib} 
\usepackage[version=3]{mhchem}
\usepackage[left=1.5cm, right=1.5cm, top=1.785cm, bottom=2.0cm]{geometry}
\usepackage{balance}
\usepackage{mathptmx}
\usepackage{authblk} 
\usepackage{sectsty}
\usepackage{graphicx} 
\usepackage{lastpage}
\usepackage{amsmath}
\usepackage{bm} 
\usepackage[format=plain,justification=justified,singlelinecheck=false,font={stretch=1.125,small,sf},labelfont=bf,labelsep=space]{caption}
\usepackage{float}
\usepackage{fancyhdr}
\usepackage{fnpos}
\usepackage[english]{babel}
\addto{\captionsenglish}{%
  
}
\usepackage{array}
\usepackage{droidsans}
\usepackage{charter}
\usepackage[T1]{fontenc}
\usepackage[usenames,dvipsnames]{xcolor}
\usepackage{setspace}
\usepackage[compact]{titlesec}
\usepackage{hyperref}

\usepackage{epstopdf}

\definecolor{cream}{RGB}{222,217,201}

\title{\LARGE\bfseries\noindent
\textit{E.~coli} bacterium tumbling in bulk and close to surfaces: A simulation study}

\author[1]{Pierre Martin}
\author[2]{Tapan Chandra Adhyapak}
\author[1]{Holger Stark}

\affil[1]{Institute of Physics and Astronomy, Theory Division, Technische Universität Berlin, Hardenbergstrasse 36, 10623 Berlin, Germany}
\affil[2]{Indian Institute of Science Education and Research Tirupati, Yerpedu P.O. PIN 517619, Tirupati, AP, India}

\date{}

\begin{document}
\maketitle

\begin{abstract}
    
\noindent\normalsize{
Motility is fundamental to the survival and proliferation of microorganisms. The \textit{E.\ coli} bacterium
propels itself using
a bundle of rotating helical flagella. If one flagellum reverses its rotational direction, it leaves the bundle, performs a
polymorphic transformation, and the bacterium tumbles.
The \textit{E.\ coli} bacterium is hydrodynamically attracted to surfaces. This prolongs its residence time, while tumbling facilitates
surface detachment.
We
develop a  
model of \textit{E.\ coli}
that uses an extended Kirchhoff-rod theory to implement
flagellar flexibility 
as well as different polymorphic conformations
and perform hydrodynamic simulations with the method of
multi-particle collision dynamics (MPCD). 
To establish a reference case, we determine the distribution of tumble angles in the bulk fluid, which shows good
agreement with experiments  for a fixed tumble time. Increasing the hook stiffness, narrows the tumble angle distribution 
and reduces the flagellar dispersion during tumbling. Close to a bounding surface, the tumble angle distribution is shifted to
smaller angles, while flagellar dispersion is reduced. Reorientation within the plane favors the forward direction, which might 
be an explanation for prolonged run times observed in experiments.}
\end{abstract}





\section{Introduction}

Bacterial motility is a prerequisite for the survivability and proliferation of bacteria.\cite{Ottemann1997, Pratt1998}
Since the first direct observation of microorganisms,\cite{Gest2004}
researchers have revealed numerous means of locomotion that bacteria use to swim in a fluid environment at low Reynolds numbers.\cite{Wadhwa2022, Lauga2016, Ishikawa2024}
In a real habitat such as the human body but also in a medical catheter or a water pipe, bacteria 
are not in their planktonic state but they encounter surfaces, where they attach to.\cite{Bray2000}
This constitutes the onset of the formation of a microcolony, which ultimately develops into
complex structures such as a biofilm.\cite{Watnick2000}
The \textit{Escherichia coli} (\textit{E.\ coli}) bacterium, the most studied microorganism, uses nano-sized rotary motors embedded in its cell wall to propel relatively rigid filaments of helical shape called flagella.\cite{Berg1993, Berg2003} During a run phase of the bacterium, all flagella rotate in the same direction and form a single tightly packed bundle.\cite{Turner2000, Berg2004, Lauga2016} To reorient itself, \textit{E.\ coli} initiates a tumble event,\cite{Darnton2007} which serves as the key mechanism for \textit{E.\ coli} to explore its environment, seek nutrients, or find a surface to attach to and proliferate.\cite{Berg1972, Costerton1995, Rob2005}
When approaching a surface, the pusher flow field generated by the bacterium attracts, aligns, and ultimately traps $E.\ coli$ at the surface.\cite{Berke2008} A direct consequence is the long residence time of bacteria at surfaces, which facilitates their attachment and the subsequent formation of biofilms.\cite{Li2011, Knut2011} 
As 
tumbling
is also a means for \textit{E.\ coli} to escape the surface, unraveling how \textit{E.\ coli} tumbles near surfaces is crucial for understanding bacterial adhesion and could help to develop strategies to mitigate biofilm formation.

A tumble event occurs when one or more flagellar motors reverse their rotation from counterclockwise to clockwise.\cite{Turner2000, Darnton2007}
This reversal causes the involved flagella to leave the bundle and undergo a series of polymorphic transitions, starting from the characteristic normal helix shape and transitioning to the final curly-I state.\cite{Turner2000,Vogel2013,Tapan2016} Once the flagella resume counterclockwise rotation, they transform back to their normal swimming conformation and the bundle reforms.
While up to twelve stable polymorphic forms have been predicted,\cite{Calladine1975} the semi-coiled and curly-I shapes are predominantly observed during tumbling.\cite{Turner2000, Darnton2007} The rotation-induced polymorphic conformations alter key helical properties 
of the flagella, such as pitch, radius, and handedness.\cite{Vogel2013,Tapan2016}
These changes modify the steric forces between the flagella as well as the flow fields generated by them, which 
strongly influences the tumbling process.\cite{Tapan2015, Tapan2016, Lee2018}

An important part of the bacterial propulsion apparatus is the flagellar hook, a $50 - 100\,\text{nm}$ 
filament that connects the shaft of the bacterial flagellar motor to the flagellum.\cite{Riley2018, Son2013, Hirano1994}
The hook is a universal joint that efficiently transfers the motor torque to the flagellum regardless of its orientation.\cite{Kato2019} 
The high flexibility of the hook is crucial for the formation of the flagellar bundle\cite{Brown2012, Bianchi2023} and its length is tightly 
regulated by biomolecular mechanisms. Ref. \citenum{Spoering2018} demonstrated that an optimal length is essential for motility. While 
a shorter hook prevents the formation of the flagellar bundle, a longer hook causes instabilities in the bundle.
Although the role of hook flexibility during tumbling is not fully understood, significant variations in its stiffness under rotation have 
been reported. Ref. \citenum{Nord2022} finds that the bending stiffness of the hook changes dynamically during rotation, while ref. \citenum{Zhang2023} shows that it increases under clockwise rotation. 

A known characteristic of pusher swimmers is their long residence time close to surfaces.\cite{Knut2011, Schaar2015, Mousavi2020} 
Moreover, chiral microswimmers
like \textit{E.\ coli} perform circular trajectories in the presence of a wall, which limits their exploration capability.\cite{Lauga2006, Perez2019, Bianchi2017} Tumbling is a key mechanism for the bacterium to escape the surface but also to change swimming direction within the surface. Several experimental studies have investigated the role of tumbling near a flat surface.\cite{Molaei2014, Lemelle2020, Junot2022} 
Ref. \citenum{Molaei2014} 
observed that tumble events mostly reorient bacteria parallel to the wall. In particular,
they found that tumbling near a surface causes smaller reorientation angles and occurs at a reduced frequency. So they concluded 
that tumbling is not very efficient to escape trapping at the surface.
In contrast, ref. \citenum{Junot2022} 
concluded that tumbling is ``a quite efficient means to escape from surfaces'' by stressing the large behavioral variability.
On the theoretical side, several
numerical models have been developed to investigate the swimming \textit{E.\ coli} close to
surfaces.\cite{Giacche2010, Mousavi2020, Hu2015_surface} 
Furthermore,
the few studies that investigated tumbling were all
conducted in a bulk fluid.\cite{Tapan2016, Kong2015, Wu-Zhang2025, Dvoriashyna2021}

To elucidate the specific role of the surface during tumbling, in this article we develop
a numerical model of \textit{E.\ coli} that captures the key features of bacterial motility, including flexible flagella and a full dynamic description of their polymorphism following our earlier work.\cite{Vogel2013, Tapan2016} 
For this, we use an extended version of the Kirchhoff-rod theory to implement a discrete model for the flexible filament, which has been proven sufficient to capture rotation-induced polymorphic transformations.\cite{Vogel2013, Tapan2016, Lee2018} The hook is not explicitly modeled, but its universal joint property and high flexibility are preserved. 
To determine the full hydrodynamic flow field initiated by the moving \textit{E.\ coli}, we employ the widely used method of multi-particle collision dynamics (MPCD). 
It has extensively been used over the past two decades and has proven to be successful in resolving the hydrodynamics of microswimmers.\cite{Zoettl2018, Hu2015, Babu2012, Zantop2020, EisensteckenThomas2016}

In this article, we use the model \textit{E.\ coli} bacterium to investigate tumbling close to a surface. First, we establish the reference case
in the bulk fluid by determining the distribution of tumble angles either for a fixed tumble time or drawn from a Gamma distribution. In particular,
for a fixed tumble time good agreement with experiments of Berg and Brown \cite{Berg1972} are achieved. Increasing the hook stiffness,
narrows the tumble angle distribution and reduces the flagellar dispersion as we demonstrate by the smallest eigenvalue of the moment of 
inertia tensor. Close to a bounding surface, we observe that the tumble angle distribution is shifted to smaller angles and reorientation
within the plane shows a strong tendency for forward-directed tumbling. Also the flagellar dispersion is reduced.

In Section\ \ref{sec:system} we introduce the model \textit{E.\ coli} bacterium and shortly explain the method of multi-particle collision 
dynamics. In Sections\ \ref{sec:Tumbling_in_bulk} and \ref{sec.surface} we present our simulation results of the tumbling \textit{E.\ coli} 
first in the bulk fluid and then close to a surface. We also discuss these results and close with conclusions in Section\ \ref{sec.conclusions}.

\section{System and simulation method}
\label{sec:system}

The \textit{E.\ coli} bacterium consists of several flagella attached to the cell body. In Section\ \ref{sec:kirchhoff} we 
first describe how we model the flagellum using  an extended version of the Kirchhoff-rod theory
and then in Section\ \ref{sec:Phy_struct} address the cell body including the motor torque driving a flagellum.
The method of multi-particle collision dynamics, which we use to perform the fluid simulations, is introduced
in Section\ \ref{subsec.MPCD}. During the course of explaining our system, we will already mention typical values 
of our parameters in real units. Then, in Section\ \ref{subsec.parameters} we will formulate them in MPCD units to 
be used in our simulations.

\subsection{Kirchhoff-rod theory to model the flagellum elasticity}
\label{sec:kirchhoff}
The bacterial flagellum is treated as a slender body described by the center-line $\mathbf{r}(s)$ where $s$ is the contour length.
We apply the Kirchhoff-rod theory to model its bend and twist elasticity.\cite{Landau1976, love2013}
The \textit{Frenet-Serret} frame $\{\mathbf{e}_1(s), \mathbf{e}_2(s), \mathbf{e}_3(s)\}$,
which forms a local orthonormal basis at location $s$, is used to quantify the deformations of the helix.
The vector $\mathbf{e}_3(s)$ represents the local tangent and $\mathbf{e}_1(s)$ and $\mathbf{e}_2(s)$ the cross section of the flagellum. Now, one introduces the rotational strain vector $\bm{\Omega}$ that transports the material tripod along the centerline, 
\begin{equation}
\partial_s \mathbf{e}_\nu = \bm{\Omega} \times \mathbf{e}_\nu, \enspace \nu =1,2,3 \, ,
\end{equation}
where $\partial_s$ means partial derivative with respect to $s$. 
The vector $\bm{\Omega}$ fully characterizes the geometry of the space curve, including the rotation of the cross section about the tangent vector $\mathbf{e}_3(s)$, which represents a twist deformation.\cite{Vogel2010} 
In particular, the bacterial flagellum can assume different helical shapes known as polymorphic configurations. They are 
defined by the ground-state vectors $\bm{\Omega}^{(n)}$.

A local deformation of the flagellum relative to $\bm{\Omega}^{(n)}$ is described by $\bm{\Omega}$. The Kirchhoff free-energy density for the elastic deformation is formulated as a harmonic approximation in $d\bm{\Omega} = \bm{\Omega} - \bm{\Omega}^{(n)}$ \cite{Landau1976},
\begin{equation}
f_K(\bm{\Omega},\bm{\Omega}^{(n)}) = \frac{A}{2}[(d\Omega_{1})^2 + (d\Omega_{2})^2] + \frac{C}{2}(d\Omega_{3})^2 \, .
\label{eq.Kirchhoff}
\end{equation}
Here, $A$ and $C$ are, respectively, the bending and twisting rigidity of the flagellum. 

A straightforward generalization of eqn.\ (\ref{eq.Kirchhoff}) to include the different polymorphic configurations can be achieved by assigning a ground state energy $\delta^{(n)}$ to each polymorphic form $\bm{\Omega}^{(n)}$. Then, the extended version of the Kirchhoff elastic free-energy density is written as \cite{Vogel2012}
\begin{equation}
f_{EK} = \min_{\forall n} \left[ f_{K}\left(\bm{\Omega},\bm{\Omega}^{(n)}\right) + \delta_{n} \right] + \frac{A}{2}\xi^{2}(\partial_{s}\bm{\Omega})^{2}
\, .
\label{eq:extended_K}
\end{equation}
The first term on the right-hand side of eqn.\ (\ref{eq:extended_K}) implies that for any rotational strain vector $\bm{\Omega}(s)$ at a point $s$, we choose the ground state vector $\bm{\Omega}^{(n)}$ of the lowest energy. As we consider local changes in the elastic energy, the transition between the polymorphic states is smooth and continuous along the flagellum, similarly to what is observed during a tumble event. This is achieved by the last term, which allows a smooth transition between two polymorphic regions of size $\xi$.\cite{Goldstein2000}.

 \begin{figure} 
\centering
  \includegraphics[width=1\columnwidth]{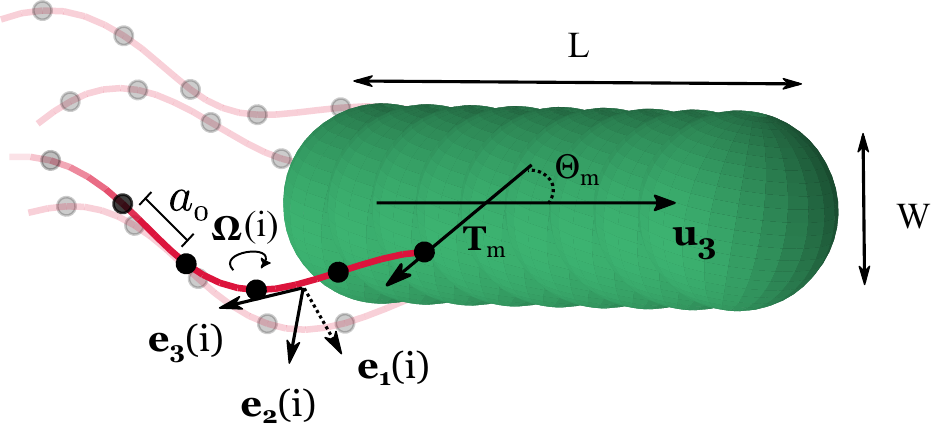}
  \caption{A schematic of the model bacterium used in our simulations.
  The cell body has a length $L$ and width $W$ as well as
  four flagella symmetrically attached at its rear. 
  The motor torque $T_m$, tilted by an angle $\Theta_m$against the body orientation $\mathbf{u}_3$,
  drives the motor tripod $\{\mathbf{e_1}(0), \mathbf{e_2}(0), \mathbf{e_3}(0)\}$ (not shown)
  and thereby it is transmitted to all the tripods
  $\{\mathbf{e_1}(i), \mathbf{e_2}(i), \mathbf{e_3}(i)\}$ along the 
  flagellum. The rotational strain vector $\boldsymbol{\Omega}^{(i)}$ transporting tripod $i$ to tripod $i+1$ is also indicated.
}
  \label{fgr:Ecoli}
\end{figure}

In our model, we allow four polymorphic states involved in a bacterial tumble event, namely normal, coiled, semi-coiled, and curly-I.\cite{Turner2000, Vogel2012, Tapan2016} 
The respective ground state vectors $\bm{\Omega}^{(n)}$ with components in the local tripod
are (in $\mu$m$^{-1}$): $\{0.00,\hspace{0.06cm}1.30,\hspace{0.06cm}-2.11\}$, $\{-0.51,\hspace{0.06cm}1.74,\hspace{0.06cm}-0.56\}$, 
$\{-1.18, \hspace{0.06cm}1.84,\hspace{0.06cm}0.98\}$, and $\{-1.80,\hspace{0.06cm}1.56, \hspace{0.06cm}2.53\}$.\cite{Vogel2012} 
While \textit{E.coli} swims in its natural environment, it exhibits the normal configuration. 
The mechanical and hydrodynamic stresses induced by the reverse rotation of the flagellum 
cause the polymorphic transitions. Multiple polymorphic states are observed in experiments, where the curly-I helix frequently 
appears as the final polymorphic shape.\cite{Turner2000, Darnton2007, Mears2014}
Therefore, we tune the unknown values of the ground state energy $\delta^{(n)}$,
which we treat as pure phenomenological parameters,
to specifically observe the curly-I state.
Studies how the choice of the ground energies influence the occurence of different polymorphic shapes were performed in
refs.\ \citenum{Tapan2016} and \citenum{Vogel2013}.
Since the reversely rotating flagellum leaves the flagellar bundle, the cell body can no longer swim on a straight path. It tumbles and
thereby changes its direction.
Finally, to maintain a constant length of the filament, a stretching free energy density is added: $f_{st} = K(\partial_{s}\mathbf{r})^2/2$. 
Then, the total elastic free energy becomes $\mathcal{F}[\mathbf{r}(s), \Phi(s)] = \int_{C_l} ds(f_{EK} + f_{st})$,
where the twist angle $\Phi(s)$ describes a rotation about the local tangent vector $\mathbf{e}_3$. 

From this elastic free energy, one can calculate the elastic forces and torques
\begin{equation}
    \mathbf{F}_{el} = -\frac{\delta \mathcal{F}}{\delta \mathbf{r}} \quad \text{and} \quad T_{el} = -\frac{\delta \mathcal{F}}{\delta \Phi} \,
\end{equation}
acting locally on the flagellum. A typical length of a flagellum is $C_l = 8 \mu\mathrm{m}$.
To resolve numerically the flagellar dynamics, the centerline
$\mathbf{r}(s)$ is discretized into $N=80$ beads of mass $M=10m$ identified by the position $\mathbf{r}_i$, where $m$ is
the mass of a single MPCD fluid particle, which we will describe in Section\ \ref{subsec.MPCD}. $N-1$ segments connect the beads and carry a \textit{Frenet-Serret} frame $\{\mathbf{e}_1(i), \mathbf{e}_2(i), \mathbf{e}_3(i)\}$, the discretized rotational strain vector $\bm{\Omega}_i$ describes the rotation of this frame attached to the $i^{th}$ segment.
To achieve a suitable hydrodynamic coupling between the flagellum and the fluid, the beads of the flagellum participate in the collision step by exchanging momentum with the solvent, as described in more detail in Section\ \ref{subsec.MPCD}.

 \subsection{Physical structure of \textit{E.coli}} \label{sec:Phy_struct}

The cell body is constructed using nine overlapping spheres so that they approximately form a spherocylinder. It is treated  
as a rigid body with a moment of inertia tensor $I_{cell}$ and a density that matches the MPCD fluid, 
so the cell body is neutrally buoyant. This model has already been demonstrated to be suitable for hydrodynamic studies and is detailed in ref. \citenum{Zantop2020}. Standard dimensions for an $E.\ coli$ cell body are a width of $W = 0.9 \mu\mathrm{m}$ and a length of 
$L = 2.5 \mu\mathrm{m}$.\cite{Berg2004} Four filaments are symmetrically anchored at the rear of the cell body, as illustrated in
Fig.\ \ref{fgr:Ecoli}. This design assumes a fully axisymmetric bacterium. The simplification allows us to perform systematic studies 
without randomly choosing the specific position of the reversely rotating flagellum to initiate tumble events. Thereby, the parameter 
space is considerably reduced. The cell body is equipped with an orthonormal basis, $\{\mathbf{u}_1, \mathbf{u}_2, \mathbf{u}_3\}$, 
which fully describes its orientation and rotation in space. 

A motor embedded in the cell wall of an $E.\ coli$ applies a constant torque, which is transmitted through the hook to a flagellum. 
A motor tripod $\{\mathbf{e}_1(0), \mathbf{e}_2(0), \mathbf{e}_3(0)\}$ with fixed direction of $\mathbf{e}_3(0)$ is placed on the surface of the cell body, so only the rotation around ${\mathbf{e}_3(0)}$ is allowed.
While our simulations cannot numerically resolve the hook, its key physical features are preserved. The motor torque $\mathbf{T}_m$ is constant and applied along the fixed direction $\mathbf{e}_3(0)$ to the flagellum. A counter torque $\mathbf{-T}_m$ is applied to the cell body to satisfy the torque-free condition. The vector $\mathbf{e}_3(0)$ is tilted by an angle $\Theta_m = 55^\circ$ 
with respect to the bacterium's long axis
$\mathbf{u}_3$, as detailed
in Fig. \ref{fgr:Ecoli}.\cite{Tapan2015} This tilt results in a counter-rotation of the cell body at a rate approximately five times slower than that of the bundle, in agreement with experiments.\cite{Darnton2007}
The orthonormal tripod, $\{\mathbf{e}_1(0), \mathbf{e}_2(0), \mathbf{e}_3(0)\}$, is coupled with the rest of the flagellum through the Kirchhoff elastic energy density as described in Section\ \ref{sec:kirchhoff}, but with reduced bending and twisting rigidity.\cite{Namba1997} As a result, the motor torque is efficiently transmitted to the flagellum, which can freely bend around the cell body, replicating the physical characteristics of the hook. 

To prevent interpenetration between the flagella and the cell body or a bounding surface, steric repulsions are implemented using the WCA potential, an adaptation of the Lennard-Jones potential:
\begin{equation}
U_{WCA =}
\begin{cases} 
4\epsilon \left[ \left(\frac{\sigma}{r}\right)^{12} - \left(\frac{\sigma}{r}\right)^6 \right] + \epsilon, & r
\leq \sqrt[6]{2} \sigma, \\
0, & \text{otherwise.}
\end{cases}
\label{eq:LJ}
\end{equation}
Here, $r$ represents the smallest distance between the flagellum and the nearby surface, while at and beyond the cutoff radius $\sqrt[6]{2} \sigma$ the interaction force is zero.
We use $U_{WCA}$ for both flagellum-flagellum and flagellum-cell body interactions, as well as when a bacterium interacts with a wall. For the first case, we have implemented a method so that flagella cannot cross each other as detailed in ref. \citenum{Tapan2015}, which also uses $U_{WCA}$.

\subsection{Fluid simulation: Multi-particle collision dynamics}
\label{subsec.MPCD}
The hydrodynamic flow fields initiated by the swimming $E.\ coli$ are determined using the method of multi-particle collision dynamics (MPCD), 
a coarse-grained particle-based solver of the Navier-Stokes equations.\cite{Malevanets1999, Malevanets2000, Gompper2009book, Zoettl2018} 
The MPCD fluid consists of point-like particles with mass $m$, and their dynamics
is resolved through a two-step procedure: the streaming step and the collision step. In the streaming step, each particle performs ballistic motion during time $\Delta t$ starting from its position $\mathbf{r}_i(t)$ with its velocity $\mathbf{v}_i(t)$:
\begin{equation}
\mathbf{r}_i(t+\Delta t) = \mathbf{r}_i(t) + \mathbf{v}_i(t)\Delta t \, .
\end{equation}

In most molecular dynamic simulations, $\Delta t$ determines the accuracy of the simulation. However, in MPCD $\Delta t$ is a parameter of the simulation that determines the fluid properties such as the viscosity.
In MPCD, particles do not interact directly with each other.
Instead, they artificially exchange momentum during the collision step
using a collision rule that preserves momentum. In practice, this is realized
by sorting particles into unit cells of a cubic lattice with lattice constant $a_0$. Within a cell $\xi$, the new velocity $\mathbf{V}_{i}$ of particle $i$ is calculated according to the special collision rule MPCD AT+a:\cite{Noguchi_2007}
\begin{equation} 
\mathbf{V}_{i} = \mathbf{u}_{\xi} + \mathbf{v}_{i}^{r} - \frac{\sum_{j=1}^{N_{\xi}}m_{j}\mathbf{v}_{j}^{r}}{\sum_{j=1}^{N_{\xi}}m_{j}} + \left[\mathbf{I}_{\xi}^{-1} \sum_{j=1}^{N_{\xi}}m_{j}\mathbf{r}_{j}^{s}\times(\mathbf{v}_{j}-\mathbf{v}_{j}^{r})\right] \times \mathbf{r}_i^s \, .
\label{eq:collision_step}
\end{equation}
Here, the velocity $\mathbf{u}_{\xi}$ represents the mean velocity of  fluid particles in
cell $\xi$, and $\mathbf{v}_{i}^{r}$ is a random velocity drawn from a Maxwell-Boltzmann distribution at temperature $T$. The third term is a correction ensuring linear-momentum conservation. The quantity $\mathbf{I}_{\xi}$ in the last term is the moment of inertia tensor calculated from the positions of all particles within the cell, and $\mathbf{r}^{s}_i$ is the relative position of a particle with respect to the cell's center of mass. The last term is needed to ensure angular momentum conservation, which proves to be essential for preserving correct physical behavior.\cite{Goetze2007}
By definition of the collision rule, the temperature remains constant and thermal fluctuations are naturally included. Finally, to
ensure Galilean invariance, the grid is randomly shifted at each MPCD step by a vector $\mathbf{s} = \{s_1, s_2, s_3\}$, where
the components $s_i$ are drawn from a uniform distribution, $s_i \in \{-a/2, a/2\}$.\cite{Ihle2001}

When a cubic cell overlaps with the cell body or a wall, virtual particles are introduced within this cubic cell to maintain the average particle 
density consistent with the surrounding fluid.\cite{Gompper2009book}
The virtual particles are important to enhance the accuracy of the hydrodynamic flow field. In particular, they improve the no-slip 
boundary condition.
The mass notation $m_j$ in eqn.\ (\ref{eq:collision_step}) highlights the distinct mass contributions of the flagellar beads in the collision step. 

To couple the bacterium with the solvent, momentum exchange is necessary. Hydrodynamic coupling between the cell body and the fluid is achieved as particles collide with the cell body. They obey the bounce-back rule that reverses the velocities of the colliding particles and thereby
enforces the no-slip boundary condition.\cite{Zoettl2018}
After the collision, the linear and angular momentum of the cell body are updated by the momenta transferred from the colliding particles 
to ensure momentum conservation.
Similarly, to satisfy the no-slip boundary condition at bounding
walls, fluid particles colliding with a wall follow the same bounce-back rule.

\subsection{Parameters of the system}
\label{subsec.parameters}
All system parameters are expressed in terms of MPCD units, for which we take the lattice constant $a_0$ as characteristic length, 
the mass $m_0$ of the fluid particles, and thermal energy $k_BT$. Using these parameters, the MPCD time unit
is defined as $\tau_0 = a_0 \sqrt{m_0 / k_BT}$, the viscosity unit as $\eta_0=\sqrt{m_0k_BT}/a^2_0$, and the resulting velocity unit becomes $v_0 = \sqrt{k_BT/m_0}$. In our simulations, the density of MPCD particles is set to 
$\rho = 10 / a_0^3$ and the MPCD time step is $\Delta t = 0.05 \tau_0$.
These quantities together with the specific collision rule determine the viscosity, which becomes $\eta = 7.5 \eta_0$.\cite{Noguchi2008}
The thermal energy is $k_BT = 4.13 \, \mathrm{pN} \, \mathrm{nm}$ at room temperature $T= 300\mathrm{K}$. 

To have a good balance between computational
efficiency and good resolution of our model bacterium, we match its dimension with $E.coli$ and set the lattice constant to
$a_0 = 0.1 \, \mu\mathrm{m}$.
By comparing with the real values mentioned in the previous sections, we can then fix all lengths in units of $a_0$. During a typical simulation of a tumble event, the bacterium is immersed in a cubic box with periodic boundaries of size $L_b=150 a_0$, approximately twice the size of the modeled bacterium and
sufficiently large to mitigate reflections of the flow field at the boundaries.
The cell body has a width and length given by $W = 9a_0$ and $L = 25a_0$, respectively. The length of a flagellum is set to $80a_0$ and the number of 80 beads are chosen such that in each cell one has a bead with mass $M=10m_0$, which ensures
matching density with the solvent.\cite{Gompper2009book, Ripoll2005}
Therefore, in the simulations the segments connecting the beads have a length of $a_0$.
For the width of the flagellum, which appears as $\sigma$ in the WCA potential of eqn.\ (\ref{eq:LJ}) for flagellum-flagellum and 
flagellum-cell body interactions, we choose $\sigma = 0.4a_0$, roughly twice the real width. When the bacterium interacts with a wall, we use $\sigma = a_0$. Finally, the energy unit in the WCA potential is set to $\epsilon =  k_BT$.

The other physical parameters of the flagellum are also chosen to be consistent with experimental observations. The magnitude of the motor torque applied to the flagellum is constant and set to $| \mathbf{T}_m |= 820 \, k_BT$,
which leads
to an average velocity $v \approx  0.03 v_0$. \cite{Darnton2007,Vogel2013, Tapan2015}

The bending and twisting rigidities of the flagellum are, respectively, $A=13310 \, k_BT a_0$ and $C=8450 \, k_B T a_0$. \cite{Darnton2007, DARNTON2007_F_extension} To prevent elongation of the flagellum, a high stretching rigidity, $K=10^5 k_BT /a_0$ is applied. As discussed in Section\ \ref{sec:Phy_struct}, the values of the ground state energies are fixed and set to
$ \delta^{(n)} = \{0.0,\hspace{0.06cm}25,\hspace{0.06cm}10,\hspace{0.06cm}0.0\}$ 
in units of $k_BT$. The values are chosen such that during reverse rotation the flagellum reaches the fourth, the curly-I state.
Although the bending rigidity of the hook varies dynamically,\cite{Nord2022, Zhang2023} we simplify the model by assuming it remains constant at $A_h=2.5 k_BT a_0$ with the twisting rigidity set to $C_h=4840k_BT a_0$. \cite{Tapan2016, Sen2004}

With the chosen parameters, the time scale for vorticity diffusion is 
$\tau_d = \rho W^2 / \eta \approx 100\tau_0$,\cite{Dhont1996}
and the Reynolds number becomes $\mathrm{Re}=\rho W v/\eta \approx 0.4$, 
where $v$ is the swimming velocity.
For each simulation of a tumble event the bacterium is first left to swim for a random number of MPCD steps such that the 
time is larger than $\tau_d$. This ensures that the flow field has been fully established and also reduces correlations between successive simulations, 
when determining the distributions of tumble angles. 

Finally, mapping the viscosity $\eta$ in the MPCD simulations to that of water, results in a bundle rotation frequency 
$2 \pi \omega \approx 600 \, \mathrm{Hz}$  and a swimming velocity $v \approx 100 \, \mu \mathrm{m} \, \mathrm{s}^{-1}$. Although these values are higher than those observed in experiments,
the ratio $v / \omega r \approx 0.1$, where 
$r = 0.2 \mu \mathrm{m}$
refers to the helical radius of the normal flagellar configuration, is consistent with experimental findings.\cite{Darnton2007}

\section{Tumbling \textit{E.\ coli} in a bulk fluid}
\label{sec:Tumbling_in_bulk}

We start with describing a tumbling \textit{E.\ coli} in a bulk fluid to have a reference case for the investigations in Section\ \ref{sec.surface}, where
we consider tumbling close to a surface. In Section\ \ref{subsec:descriptionTB} we introduce a single tumble event, 
either with fixed tumble time or drawn from a Gamma distribution, and then describe a typical event in detail. Then, in 
Section\ \ref{sec:long_traj_bulk} we show a long trajectory of a running and tumbling bacterium with run and tumble times drawn from 
respective Gamma distributions. Finally, in Section\ \ref{sec:analyze_tb_event_bulk} we characterize the tumble events by discussing tumble 
angle distributions, the influence of hook bending rigidity, and the flagellar dispersion during tumbling.

\subsection{Description of a tumble event}
\label{subsec:descriptionTB}

The duration of a tumble event 
of $E.\ coli$ typically ranges between $0.1$ and $0.3 \mathrm{s}$.
\cite{Berg1972, Zijie2018, Seyrich2018} 
In our model, a tumble event is triggered by reversing the direction of the motor torque.
In the following, we take the tumble duration $\tau_t$ as the time during which the reverse torque is applied.
We express this duration in terms of the period of the bundle rotation,
$\tau_b$. The 
flagellar bundle
of an $E.\ coli$ typically rotates
at a frequency of $100 \, \text{Hz}$.\cite{Darnton2007} Thus the
tumble duration of $0.3 \mathrm{s}$,
corresponds to $30 \tau_b$.
We investigate two cases.
In the first scenario, the tumble time 
assumes the fixed value
$\tau_t = 30 \tau_b$.
In the second scenario, we allow the tumble time to vary, as observed in experiments. We assume that it follows a
Gamma distribution
\begin{equation}
\label{eq:gamma_eq}
P(\tau_t; k, \theta) = \frac{1}{\Gamma(k) \theta^{k}} \tau_t^{k-1} e^{-
\tau_t / \theta
} \quad \mathrm{with} 
\quad \Gamma(k) = \left(k-1\right)!
\end{equation}
where
$k$ and $\theta$ are the 
shape and scale parameters,
respectively. We chose $k=2$ and the 
$\theta = 15\tau_b$, which 
gives
a mean tumble time of 
$ \langle \tau_t \rangle =30 \tau_b$. In particular, the distribution grows linearly at small $\tau_t$.
The use of the Gamma distribution is motivated by 
experimental observations 
in 
ref.\ \citenum{Junot2022}.

A tumble is initiated by reversing the rotational direction of one flagellum. For this, the
torque is linearly switched in time
from $-T_m$ to $T_m$ over a short 
period
of $3.5 \tau_b$.
Once the tumbling time 
has elapsed, the torque is 
smoothly 
reversed back to $-T_m$ during time $3.5\tau_b$.

 \begin{figure} 
\centering
  \includegraphics[width=\columnwidth]{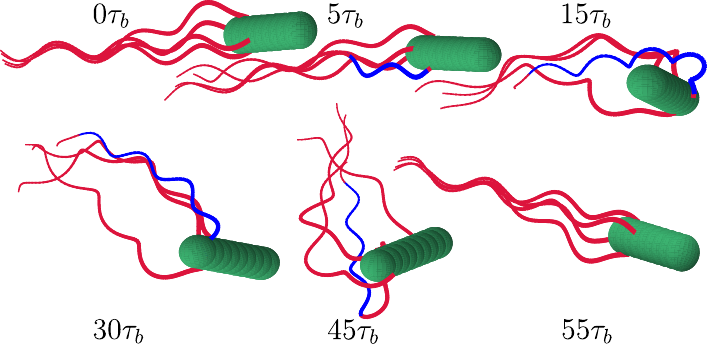}
  \caption{Simulation snapshots of a tumble event  at different time points. The flagellum shown in red represents the normal 
  helical configuration, while in the reversely rotating flagellum blue indicates the transition to the curly-I state propagating 
  along the flagellum. 
The reverse rotation 
starts
at $0 \tau_b$ and ends at $ 30 \tau_b$, as detailed in the main text.
}
  \label{fgr:Tumble_event}
\end{figure}

A typical tumble event 
with a tumble duration of $30\tau_b$
is illustrated by snapshots from our simulations in Fig.\ \ref{fgr:Tumble_event}
and supplemental video A1.
Different phases can be distinguished. The first phase lasts until $t=15 \tau_b$. It starts with initiating the reverse rotation so that the 
flagellum switches from the normal (red) to the curly-I (blue) configuration. The viscous drag on the reversely rotating flagellum initiates
a reversal of the local helical torsion so that the transition to the curly-I state rapidly propagates from the cell body to the tip of the flagellum covering 95\% of its full length. The speed and extent of this transition strongly depends on the ground state energies $\delta^{(n)}$ and the motor torque.\cite{Vogel2013, Tapan2016}
Interestingly, when initiating the transition to the curly-I state, the flagellum hardly rotates while the other flagella push the cell body forward
so that it always tilts towards the transforming flagellum (see Fig.\ \ref{fgr:Tumble_event}, snapshot at $ t = 5\tau_b$).
In the second phase, $15\tau_b< t <30 \tau_b$, we observe that in most cases the flagellar bundle disintegrates and the single
filaments buckle. The reversely rotating flagellum
either leaves the remaining part of the bundle or stays entangled. In both case, the propelling structure is largely dispersed as shown 
in Fig. \ref{fgr:Tumble_event}.
Note that our simulations deviate here from the typical representation of a tumble event that shows the reversely rotating
flagellum outside an intact bundle \cite{Darnton2007}. However, experimental observations also reveal a diverse behavior of the bundle during tumbling reminiscent of our simulations \cite{Turner2000, Turner2016, Junot2022}.
Finally, once the motor torque has reversed back, in the time period $30\tau_b < t < 55\tau_b$
the flagellum transitions from the curly-I configuration to the normal state 
and the bundle forms again.
Note that our formal definition of the tumble duration as the time between two reversals of the motor torque
does not fully reflect the duration of a tumble event, since the reformation of the bundle also takes time.

We simulated a large number of 
single tumble events.  To quantify the reorientation of the swimming $E.\ coli$ bacterium,
we defined the tumble angle $\gamma$
as
\begin{equation} 
\gamma = \arccos{\left(\hat{\mathbf{u}}_i \cdot \hat{\mathbf{u}}_f\right)} \, .
\label{eq:tb_angle}
\end{equation}
Here, the unit vector $\hat{\mathbf{u}}_i$ 
describes the swimming direction or
orientation of the bacterium before reversing the rotation of a motor torque,
while $\hat{\mathbf{u}}_f$ refers
to the orientation 
once the flagella are fully rebundled. 
To avoid strong influences from the wobbling cell body on the tumble angle,
we adopted a more robust definition of orientation by 
taking the orientation vector from
the center of mass of the bundle to that of the cell body.

\subsection{Long trajectory: An $E.\ coli$ runs and tumbles}
\label{sec:long_traj_bulk}

 \begin{figure} 
\centering
  \includegraphics[width=0.9\columnwidth]{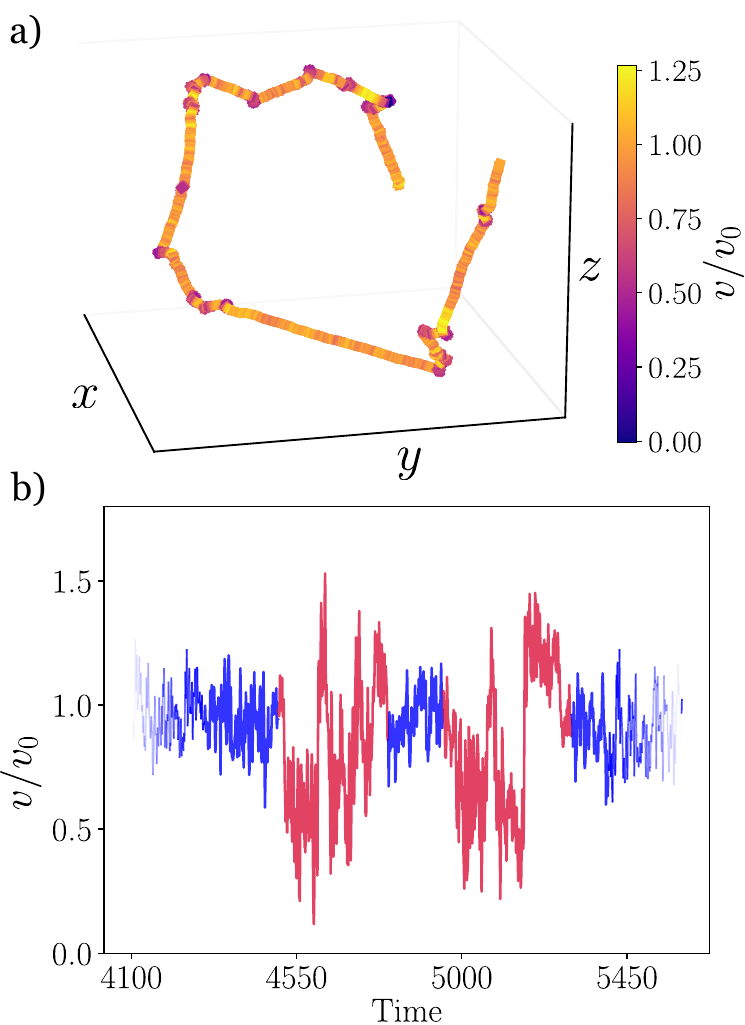}
  \caption{(a) Long simulated trajectory of the 
  $E.\ coli$ bacterium.
  Both, the run and the tumble durations 
  are taken from Gamma distributions, with respective mean run and tumble times 
  of $90\tau_b$ and $30\tau_b$.
The color map 
quantifies the velocity of the bacterium normalized by the 
swimming velocity $v_0$.  Dark blue indicates a drop of velocity
corresponding to
a tumble event.
(b) Swimming velocity of the bacterium plotted versus time taken from the trajectory in (a) for two tumble events.
The blue and red color represent the velocities during running and tumbling phase, respectively.
}
  \label{fgr:Bulk_trajectory}
\end{figure}

We also simulated long trajectories of the tumbling \textit{E.\ coli} with a total duration
of $1455 \tau_b$. The bacterium is immersed into an unbounded MPCD fluid with periodic boundary 
conditions and moves through alternating run and tumble phases. The durations of both phases follow here the
Gamma distribution of eqn.\  (\ref{eq:gamma_eq}) with respective mean run and tumble times
 of $90 \tau_b$ and $30 \tau_b$. 
Figure \ \ref{fgr:Bulk_trajectory}a)
illustrates the trajectory of the 
simulated \textit{E.\ coli} bacterium. During the run phase, all the flagellar motors rotate
in the same direction. Hydrodynamic and steric interactions synchronize the rotation of the flagella so that they form a
bundle. 
Its helical structure drives the cell body in 
a nearly straight trajectory with a constant swimming
velocity $v_0$. Small changes in the
direction 
are due to rotational diffusion.\cite{Berg2004} 

The color map in Fig. \ref{fgr:Bulk_trajectory}a) represents the swimming velocity of the bacterium
along the trajectory. We define it as
\begin{equation}
    v=\mathbf{v}_G \cdot \hat{\mathbf{u}} \,
\end{equation}
where $\mathbf{v}_G$ is the velocity of the midpoint between the centers of mass of the bundle and the cell body,
and $\hat{\mathbf{u}}$ is the orientation vector connecting both centers of mass.
Before each change in direction, the color code indicates a decrease in velocity. 
Note that towards the end of the trajectory a tumble event occurs with nearly no change in the swimming direction. 

Figure\ \ref{fgr:Bulk_trajectory}b)
provides a detailed account of the instantaneous velocity.
During the tumbling phase (indicated in red),
the velocity decreases but it also strongly fluctuates. As described in the previous subsection, 
the flagella are misaligned with the cell body and exhibit strong buckling. This deformation of the helical structure results in irregular propulsive forces acting
from the 
rotating flagella onto
the cell body, which thereby performs
erratic changes in 
its orientation as illustrated in Fig.\ \ref{fgr:Tumble_event}. 
This can also
give rise to a sudden increase of velocity beyond $v_0$, which is visible in Fig.\ \ref{fgr:Bulk_trajectory}b).
Finally, once the flagellum reverts back and the bundle 
reforms, the cell body resumes swimming at a constant velocity $v_0$. 

\subsection{Analysis of the tumble event}
\label{sec:analyze_tb_event_bulk}
 
 \begin{figure}
\centering
  \includegraphics[width=0.9\columnwidth]{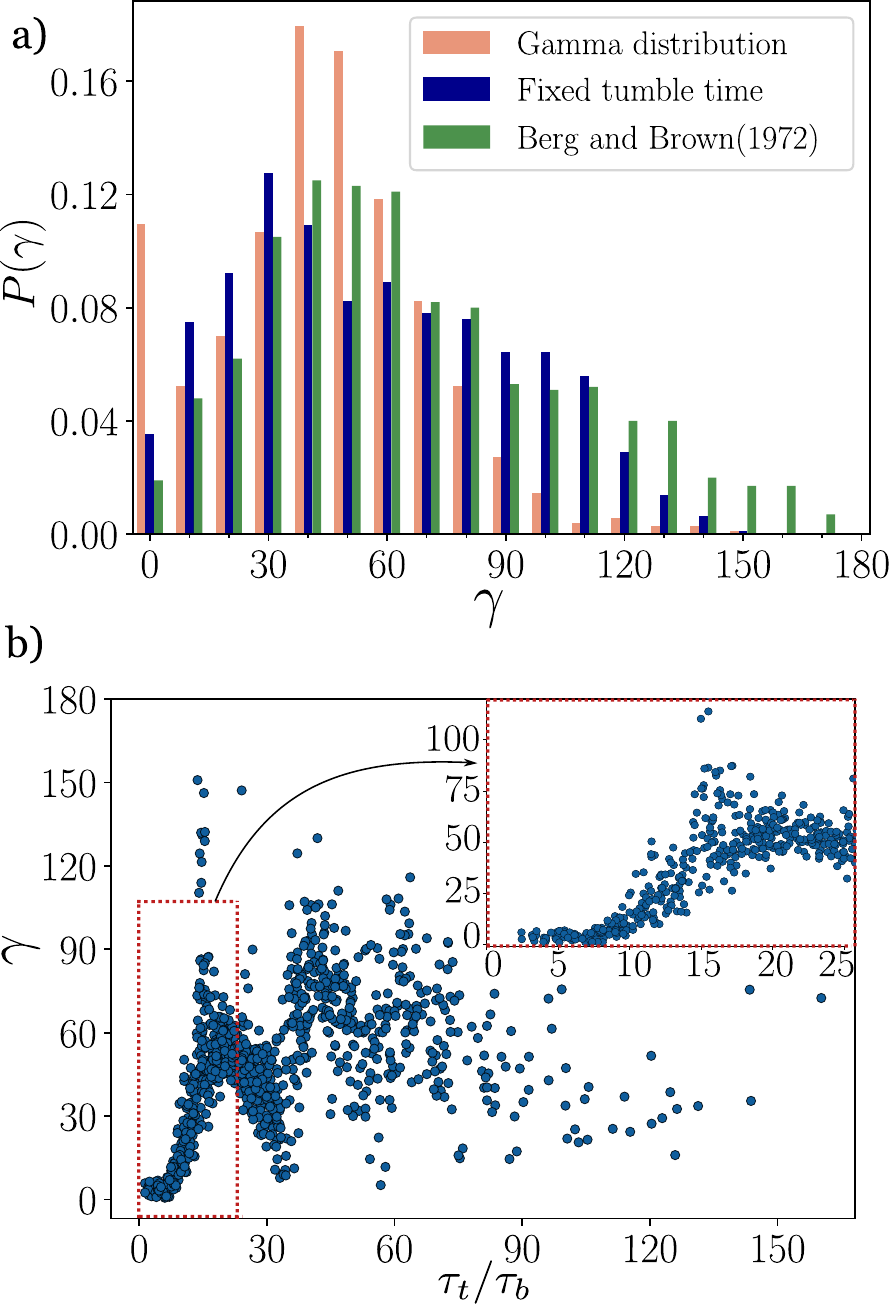}
  \caption{(a) Bar plot of the probability distribution $P(\gamma)$
  of the tumble angle $\gamma$. The green data points are reproduced from 
  ref. \citenum{Berg1972}. 
The blue and salmon data points belong to simulations, where the tumble time $\tau_t$ was either fixed or chosen from the 
 Gamma distribution $P(\tau_t;2,15\tau_b)$ of eqn.\ (\ref{eq:gamma_eq}), respectively.
(b) Tumble angle $\gamma$ plotted versus tumble time $\tau_t$ chosen from $P(\tau_t;2,15\tau_b)$.
The red frame highlights the region for small $\tau_t$ shown in the inset.
}
  \label{fgr:PDF_bulk}
\end{figure}

How much the bacterium reorients during a tumble event varies significantly both in experiments and our simulations. To quantify this, we present in Fig. \ref{fgr:PDF_bulk}a) the distribution $P(\gamma)$ of the tumble angle for simulations where the tumble time $\tau_t$ was either fixed (blue) or chosen from the Gamma distribution $P(\tau_t;2,15\tau_b)$ of eqn.\ (\ref{eq:gamma_eq}) (salmon).
The distributions were obtained from $932$ or $1032$ independent simulations, respectively.
For a fixed tumble time, $P(\gamma)$ with mean tumble angle $\langle \gamma \rangle = 61^{\circ}$ and a standard deviation of $
\Delta \gamma =34^{\circ}$ is in good agreement with the measurements of Berg and Brown in ref.\ \citenum{Berg1972}, where they
obtained $\langle \gamma \rangle = 62^{\circ}$ and $\Delta \gamma =26^{\circ}$.
 
The tumble angle distribution for tumble times chosen from a Gamma distribution
roughly reproduces the most probable tumble angle 
in the experiments but does not reach to large tumble angles. As a result the mean tumble angle of $49^{\circ}$ is smaller while $\Delta \gamma =27^{\circ}$ is comparable to the previous values.
Most prominently, a peak appears for angles between $0^{\circ}$ and $10^{\circ}$, which results from short tumble events.
This is illustrated in Fig.\ \ref{fgr:PDF_bulk}b), where we correlate tumble angle with tumble time.
The inset shows that a noticeable reorientation of \textit{E.\ coli} only occurs at times larger than $10 \tau_b$.
Prior to this 
time, the reversely rotated flagellum is not fully transformed into the
curly-I state and the bundle is still intact (see Fig. \ref{fgr:Tumble_event}).
Therefore, such tumble events,
for which we show an example in supplemental video A2,
might have been overlooked in experiments and their small tumble angles not recorded.

For tumble times $10 \tau_b < \tau_t < 15 \tau_b$, there is a nearly one-to-one relation between tumble angle and tumble time meaning the
bacterium moves deterministically. Then, around $ \tau_t \sim 15 \tau_b$ the tumble angle spreads widely and even reaches 
very large values at $ \tau_t \approx 15 \tau_b$.
This behavior is due to the hook's high flexibility, which
enables the cell body to rotate freely so that it no longer moves deterministically. 
In particular, at $ \tau_t \approx 15 \tau_b$ cell body and flagella point in opposite directions.
Finally, as the tumble time extends beyond $\tau_t \sim 30\tau_b$, the tumble angle becomes increasingly scattered,
although we always use the same attachment points for all four flagella. This shows
the complex interplay of hydrodynamic forces, flagellar elasticity, and cell body dynamics. Together, these factors give rise to a highly dynamic and intricate reorientation process. 

Recent studies have shown that the stiffness of the hook 
strongly increases with angular velocity of the motor.\cite{Nord2022, Zhang2023}
To capture the role of hook flexibility during a tumble event, we increase its bending 
rigidity $A_h$ by a factor of $10$, $100$, and $400$, respectively, such that it is still at least one order of magnitude smaller than 
the flagellar bending rigidity.  Furthermore, we choose a fixed tumble time in the simulations. As shown in refs.\ \citenum{Brown2012} and \citenum{Spoering2018}, hook stiffness plays a crucial role in the motility of \textit{E.coli}.
since an increased stiffness prevented the formation of the bundle.
In our simulations we made sure that the flagella still formed a bundle prior to a tumble event, for all the chosen rigidities.

 \begin{figure} 
\centering
  \includegraphics[width=0.95\columnwidth]{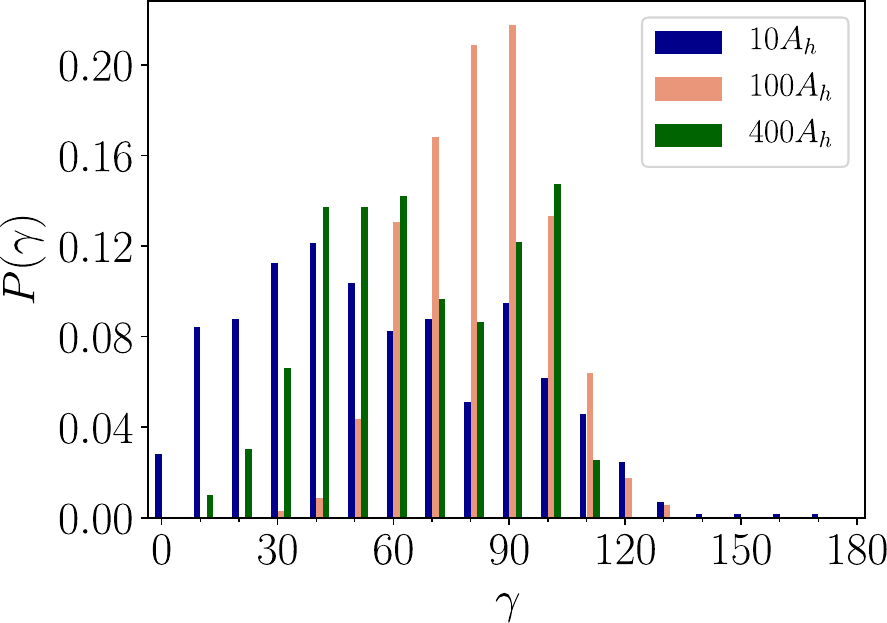}
  \caption{
    Bar plot of the probability distribution $P(\gamma)$
    for three different values of the hook bending rigidity. The fixed tumble time $\tau_t = 30 \tau_b$ is used in the simulations.
  }
  \label{fgr:PDF_bulk_hook}
\end{figure}

Fig. \ref{fgr:PDF_bulk_hook} shows
the tumble angle distribution $P(\gamma)$ 
for the three hook bending rigidities, mentioned above. 
The distributions are constructed from independent simulations, with a total of
$570$, $346$, and $624$ simulations for the respective rigidities
$10A_h$, $100A_h$, and $400A_h$.
For a hook bending rigidity of $10 A_h$ only small changes occur relative 
to the reference case $A_h$.
Therefore, we do not show the latter in Fig.\ \ref{fgr:PDF_bulk_hook}.
However, for rigidities of $100 A_h$ and $400 A_h$, 
the distributions become more narrow and only tumble angles
between $20^{\circ}$ and $ 120^{\circ}$ do mainly occur.
The respective mean tumble angles of $87^\circ$ and $69^\circ$ are above the value of $61^\circ$ for the reference case.
The supplemental videos A3 and A4 show that the increased hook bending rigidity more and more
impairs the rotation of the reversely rotating flagellum. Therefore, the friction forces are significantly reduced and the polymorphic 
transformation to the curly-I state is delayed.
For a hook rigidity of $1000 A_h$ the motor stalls completely and the polymorphic transition no longer occurs.
These observations emphasize the key role of the hook bending rigidity in the bacterial motility,
affecting directly the tumble angle distribution.
Our simulations also reveal
that 
the increased hook bending rigidity significantly reduces the dispersion of the flagellar bundle during tumbling. 

To quantify the flagellar dispersion, we calculate the moment of inertia tensor
$\mathbf{I}$ of the bundle,
\begin{equation}
    I_{ij} = m\sum_{k=1}^{N}  \left( 
  |\mathbf{r}^{(k)}|^2
 \delta_{ij} - x_i^{(k)} x_j^{(k)} \right) \, ,
\end{equation}
where $N$ is the number of beads with mass $m$
forming the four flagella, 
$\mathbf{r}^{(k)} = (x_1^{(k)}, x_2^{(k)}, x_3^{(k)})$ the position
vector of bead $k$ taken relative to the center of mass of the bundle, and $\delta_{ij}$ the Kronecker delta.
We calculate the eigenvectors and eigenvalues of $\mathbf{I}$
that provide the main rotational axes and moments of inertia of the flagellar bundle.
For an intact bundle, the smallest eigenvalue $I_1^\text{bundle}$ refers to rotations about the bundle axis. To quantify the dispersion
of the bundle during a tumble event, we use the time average of the smallest eigenvalue over a complete tumble event and denote
it with $I_1$. The tumble event
begins at the reverse rotation of the motor and ends once the flagella have fully rebundled.
Thus, a value of $I_1$ larger than $I_1^\text{bundle}$ measures the flagellar dispersion.

\begin{figure} 
\centering
  \includegraphics[width=\columnwidth]{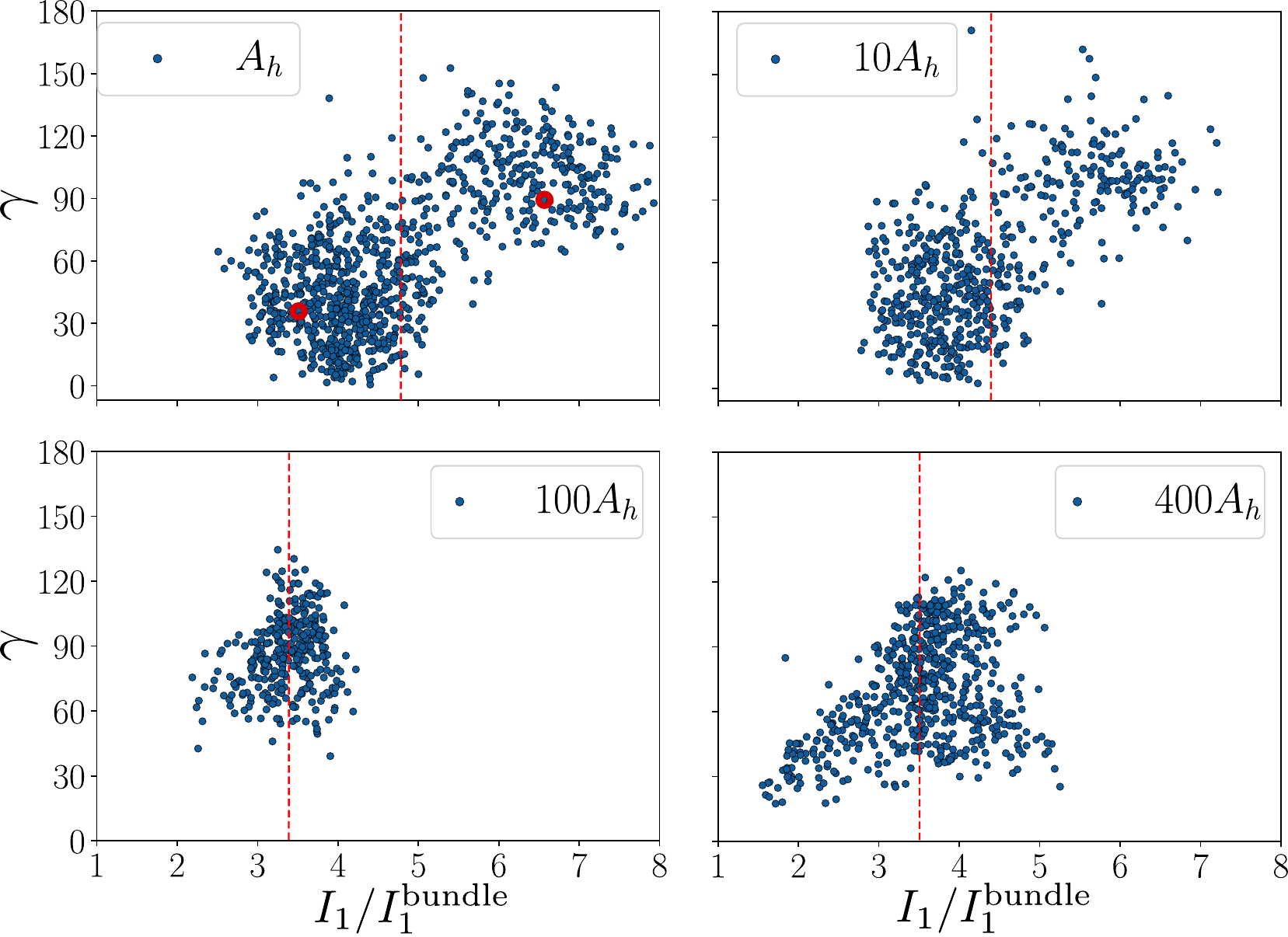}
  \caption{Scatter plots 
 of the tumble angle $\gamma$ 
 \emph{versus} time-averaged
 mo\-ment of inertia $I_1/I^\text{bundle}_1$ for 
 different
 values of hook bending rigidity. The red 
 dashed
 lines highlight the mean value 
 $\langle I_1 \rangle / I^\text{bundle}_1$. Cases shown in supplemental videos A1 and A5  are indicated by  red circles.
}
  \label{fgr:I_1_inbulk}
\end{figure}

In Fig.\ \ref{fgr:I_1_inbulk} we show scatter plots for all the tumble events that were used 
in Figs.\ \ref{fgr:PDF_bulk}a) and \ref{fgr:PDF_bulk_hook} for different hook bending rigidities. They reveal correlations 
of $I_1/I_1^\text{bundle}$ with the tumble angle $\gamma$. The dashed lines show the mean values of $I_1/I_1^\text{bundle}$
averaged over all tumble events. One immediately recognizes that increasing the hook bending rigidity significantly reduces the 
flagellar dispersion, meaning $\langle I_1 \rangle /I_1^\text{bundle} $ decreases.

For hook bending rigidities $A_h$ and $10A_h$ we discern two distinct regions. Tumble angles between $0^\circ$ and $90^\circ$ 
belong to smaller $I_1$, while tumble angles above $90^\circ$ clearly correlate with large $I_1$. We illustrate the two cases with videos
for rigidity $A_h$. Supplemental video A1, which corresponds to the tumble event illustrated in Fig.\ \ref{fgr:Tumble_event},
shows a small flagellar dispersion $I_1/I^\text{bundle}_1 = 3.5$ and a tumble angle $\gamma = 32^{\circ}$. The flagellar bundle spreads and separates into two distinct parts, with one flagellum (not necessarily reversely rotating) separated from the others.
In contrast, supplemental video A5 illustrates 
the large dispersion case with $I_1/I^\text{bundle}_1 =6.6$
and $\gamma = 92.8^{\circ}$.
Interestingly, the bundle remains tightly packed throughout the entire tumble phase but then fully unbundles once the motor of the fourth flagellum reverses back.
At this moment, the cell body is roughly antiparallel to all the flagella and to unit
all of them
in a single bundle behind the cell body, they first have to fully unbundle.

At hook bending rigidities of $100A_h$ and $400A_h$, the mean value $\langle I_1 \rangle / I_1^\text{bundle}$ is significantly reduced
and tumble angles above $120^\circ$ hardly occur. The reason is that the bundle formed by the three flagella remains nearly intact during tumbling as 
the supplemental videos A3 and A4 show.
Furthermore, as we already noted, the reversely rotating flagellum is strongly
impaired by the increased bending rigidity of the hook, which 
delays the polymorphic transition to the curly-I
state.
In contrast to the previous case, 
the larger hook rigidity
also impairs the free rotation of the cell body relative to all the flagella,
since the stiffer hook acts with a torque on the cell body. It therefore reorients less and large tumble angles do not occur.
Note, for the rigidity of $400 A_h$ the bundle stays even more compact compared to the rigidity of $100 A_h$ and the reversely 
rotating flagella protrudes more strongly from the cell body
(see supplemental videos A3 and A4).
This is very reminiscent of the typical description of a tumble event in experiments.\cite{Darnton2007}

\section{Tumbling \textit{E.\ coli} close to a flat surface}
\label{sec.surface}

This section performs the same type of analysis as described in Section \ref{sec:Tumbling_in_bulk}
but now with the \textit{E.\ coli} moving in the proximity of a
flat surface. Section \ref{sec:long_traj_surface}  presents
the simulation of an extended trajectory of the \textit{E.\ coli} using the same run and tumble characteristics as
in  Section \ref{sec:long_traj_bulk}. Then, Section \ref{subsec:analysis_tb_surface} analyzes tumble events near the surface and compares
the tumble angle probability distribution $P(\gamma)$ to the bulk case. 
It also introduces the in-plane angle $\phi$ and the polar angle $\theta$
as well as their distributions to characterize the reorientation of the \textit{E.\ coli}  
close to the surface. Finally, we relate these angles also to the dispersion of the flagellar bundle during tumbling.

\subsection{Long trajectory: An \textit{E.\ coli} 
close
to a flat surface}
\label{sec:long_traj_surface}

\begin{figure} 
\centering
 \includegraphics[width=\columnwidth]{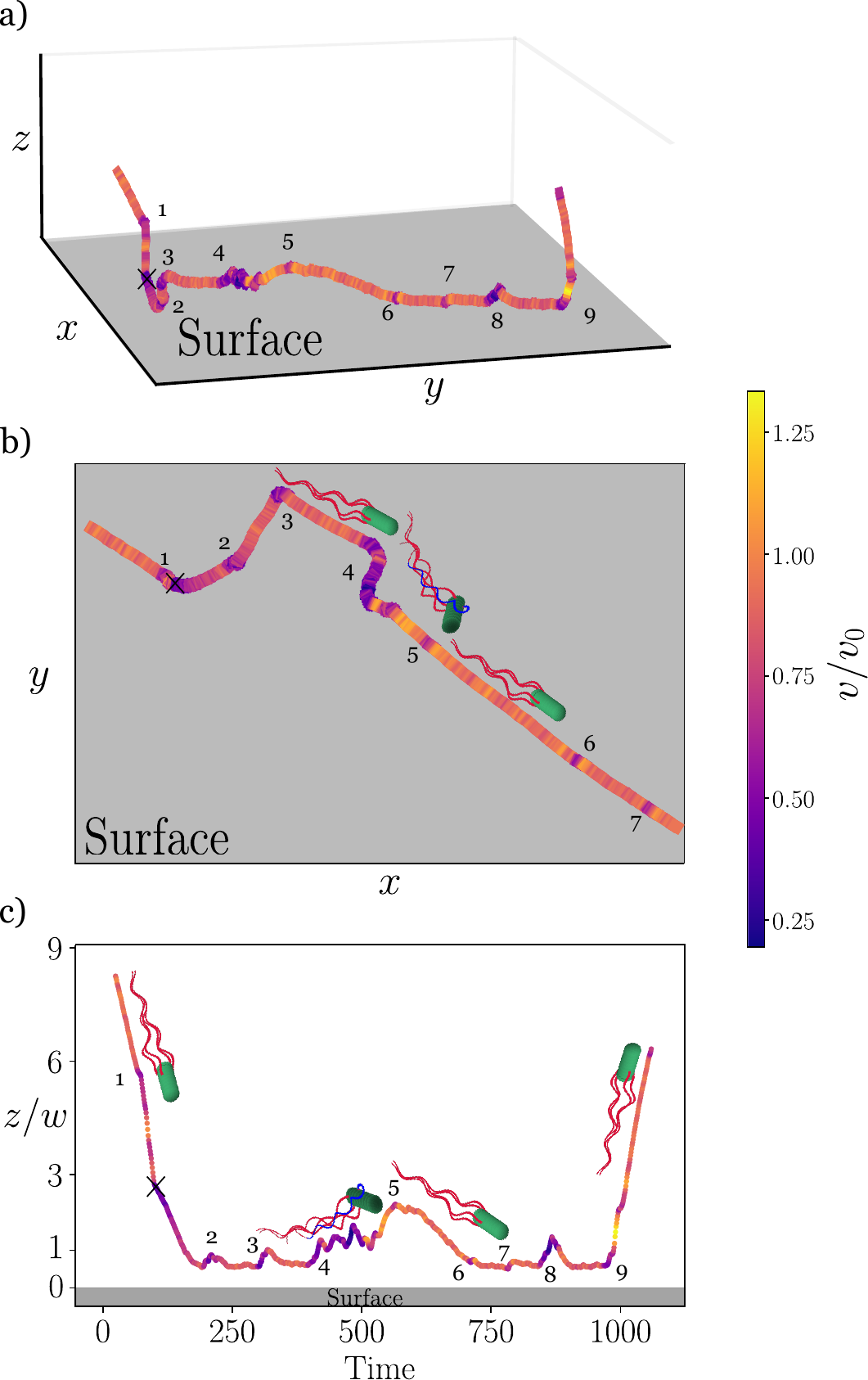}
  \caption{Run and tumble trajectory of 
  the center of mass of an
  \textit{E.\ coli} close to a surface 
  shown
  from different perspectives. The numbers enumerate and locate the successive tumble events.
  The black cross between 1 and 2 indicates where the bacterium collides with the surface.
  The color map records the change in the swimming velocity.
  (a) Three-dimensional representation of the
  trajectory starting on the left. The
  bottom surface is colored in grey to emphasize the bacterium's proximity to it. (b) Top view and (c) side view. The distance $z$
  to the surface as a function of time is
  rescaled with the width $w$
  of the cell body to highlight its proximity to the wall.
}
  \label{fgr:surface_trajectory}
\end{figure}

As a bacterium leaves the bulk fluid,
it swims towards and hits a bounding surface, \cite{Schaar2015} 
where steric as well as hydrodynamic interactions with the surface reorient the bacterium.\cite{Berke2008, Schaar2015}
Ultimately, the bacterium becomes trapped by hydrodynamic interactions \cite{Berke2008}
and follows a circular trajectory that significantly reduces its
exploration capability.\cite{Lauga2006, Giacche2010} 
Here, tumbling
serves as a key mechanism for bacteria to escape from the surface. Similar to Section \ref{sec:long_traj_bulk}, we simulated 
a long trajectory over a duration of $1400 \tau_b$, where
run and tumble times were drawn from the same Gamma distributions as in the bulk fluid.
Two parallel no-slip walls, with distance $150 a_0$ and the normal pointing along the $z$ direction, bound the fluid,
while periodic boundary conditions are applied in the 
$x$ and $y$
directions, respectively.

When we simulate a long trajectory of the bacterium, eventually, it orients towards and reaches one surface. 
This is illustrated from different perspectives in Fig.\ \ref{fgr:surface_trajectory}, where we show a segment of the long trajectory.
Along the bacterial path we record nine tumble events, each marked by a drop in swimming velocity as in the bulk fluid and in agreement with experimental observations.\cite{Lemelle2020}
Initially, the bacterium 
swims
towards the surface and undergoes one tumble event in the bulk fluid
[event 1 in Fig.\ \ref{fgr:surface_trajectory}c)].
Then, it
collides with the wall,
which we mark by a cross on the trajectory. This
shows
that hydrodynamic reorientation due to the presence of a wall 
is
not sufficient to prevent the collision,\cite{Berke2008,Schaar2015} 
which is
consistent with simulations in ref. \citenum{Mousavi2020}. 
Once near the surface, 
the bacterium
swims parallel to it and 
goes through several
tumble events 
(2 to 9 in Fig.\ \ref{fgr:surface_trajectory}). Thus, a
total of eight tumbles 
are necessary
to successfully reorient away from the wall and leave the surface.
This highlights that the \textit{E.\ coli} bacterium
preferentially reorients parallel to 
a bounding surface during tumbling. In addition, the top-down view of Fig.\ \ref{fgr:surface_trajectory}b) shows that
the in-plane reorientation 
can also
be relatively small. 
Of course, as illustrated
in Fig.\ \ref{fgr:PDF_bulk}
a small reorientation is also caused by the short duration of the reversely rotating flagellum.
However, as we show in the following, it is a specific feature of tumbling close to a surface.

\subsection{Analysis of the tumble event}
\label{subsec:analysis_tb_surface}

 \begin{figure*}
\centering
  \includegraphics[width=0.95\textwidth]{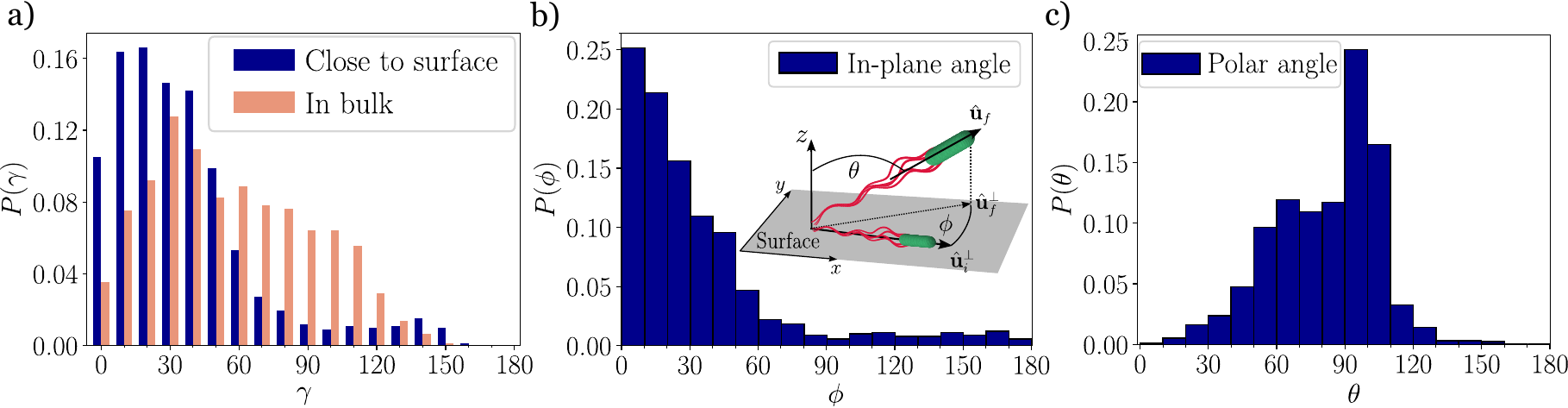}
  \caption{Bar plots of the probability distributions of different reorientation angles during tumble events: (a) Comparison between tumble angle distributions
  $P(\gamma)$ in bulk and close to a surface,
  drawn in 
  salmon and blue color, respectively.
  (b) Distribution $P(\phi)$ of the in-plane reorientation angle $\phi$. Inset: Definition of the reorientation angles $\phi$, $\theta$.
   $\hat{u}_i^{\perp}$ and $\hat{u}_f^{\perp}$ are the respective projections of the 
   initial ($\hat{u}_i$) and final ($\hat{u}_f$) orientation vectors
   on the surface.
  (c) Distribution $P(\theta)$ of the polar reorientation angle $\theta$ measured
  relative to the surface normal, 
  see inset of (b).
}
  \label{fgr:PDF_near_surface}
\end{figure*}

To further investigate
tumbling close to a surface as apparent in
Fig.\ \ref{fgr:surface_trajectory}, we perform independent simulations of single
tumble events, where we fix the reverse rotation
of one flagellum to the constant time
$\tau_t = 30 \tau_b$, as before in the bulk fluid.
For each simulation, 
the bacterium is placed close to and parallel to the surface so that it swims at a steady height \cite{Giacche2010,Mousavi2020}. 
Then, tumbling is initiated by reversing the rotation of one flagellum,
as described in Section\ \ref{subsec.parameters}.
Supplemental videos
A6 (side view) and A7 (top view) show a typical tumble event close to a surface.
Figure \ref{fgr:PDF_near_surface}a) compares the tumble angle distributions $P(\gamma)$ in the bulk 
fluid and close to the surface.
There,
the distribution is clearly strongly shifted towards 
smaller tumble angles
with a mean value $\langle \gamma \rangle = 41^{\circ}$ and a standard deviation $\Delta \gamma =32^{\circ}$. 
In the bulk fluid the respective values were
$\langle \gamma \rangle = 61^{\circ}$ and $\Delta \gamma =34^{\circ}$.

To better characterize the
orientation of the bacterium after tumbling, we calculate the 
polar ($\theta$) and in-plane ($\phi$) reorientation angles as illustrated in the inset of Fig.\ \ref{fgr:PDF_near_surface}b),
\begin{equation}
    \theta = \arccos{\left(\hat{\mathbf{z}} \cdot \hat{\mathbf{u}}_f\right)}  , \quad
    \phi = \arccos{\left(\frac{\hat{\mathbf{u}}_{i}^{\perp} \cdot \hat{\mathbf{u}}_{f}^{\perp}}{|\hat{\mathbf{u}}_{i}^{\perp}||\hat{\mathbf{u}}_{f}^{\perp}|} \right).}
\end{equation}
Here, $\hat{\mathbf{z}}$ is the unit vector
normal to the surface and $\mathbf{u}_i^{\perp}$,
$\mathbf{u}_f^{\perp}$ are the respective projections of 
the initial ($\mathbf{u}_i$) and final ($\mathbf{u}_f$) orientation vectors on the surface.

The in-plane tumble angle distribution $P(\phi)$ 
plotted
in Fig. \ref{fgr:PDF_near_surface}b)
shows a strong tendency for forward-directed tumbling with small in-plane reorientation.
In fact, $25\%$ of the recorded tumble
events show an in-plane angle smaller than $10^{\circ}$. This observation aligns with the simulated
long trajectory 
in Fig.\ \ref{fgr:surface_trajectory}b), which demonstrates that tumble events 
often result in small in-plane reorientations as illustrated by supplemental video A7. The experimental study of ref. \citenum{Molaei2014} reports a reduced frequency of tumble events near a surface and concludes
that a surface suppresses tumbling. 
In contrast,
our simulations indicate that in-plane reorientation is small [Fig. \ref{fgr:PDF_near_surface}b)]
and, therefore, such events might have been overlooked in the experiments. In fact, the trajectory in ref. \citenum{Molaei2014}
shows events with a strong decrease in velocity, which are not documented as tumbling.

The distribution of the polar reorientation angle is
shown in Fig.\ \ref{fgr:PDF_near_surface}c). 
It peaks around $\theta = 90^{\circ}$, which
is consistent with our observations 
of long trajectories at the surface
(Fig.\ \ref{fgr:surface_trajectory}) and with experimental findings \cite{Molaei2014, Junot2022}. 
The experiments show that after tumbling near a surface, bacteria tend to reorient parallel to it.
We can also deduce the escape rate from $P(\theta)$, as a polar angle $\theta$ smaller than $90^{\circ}$
means that the bacterium is oriented away from the surface.
Our analysis reveals that $53\%$ of tumble events 
result in the bacterium swimming away from the surface. But it does not necessarily fully
escape as the side view of the trajectory in Fig.\ \ref{fgr:surface_trajectory}c) illustrates.
An example of an escaping bacterium 
is shown in suppemental video A8.

 \begin{figure} 
\centering
  \includegraphics[width=\columnwidth]{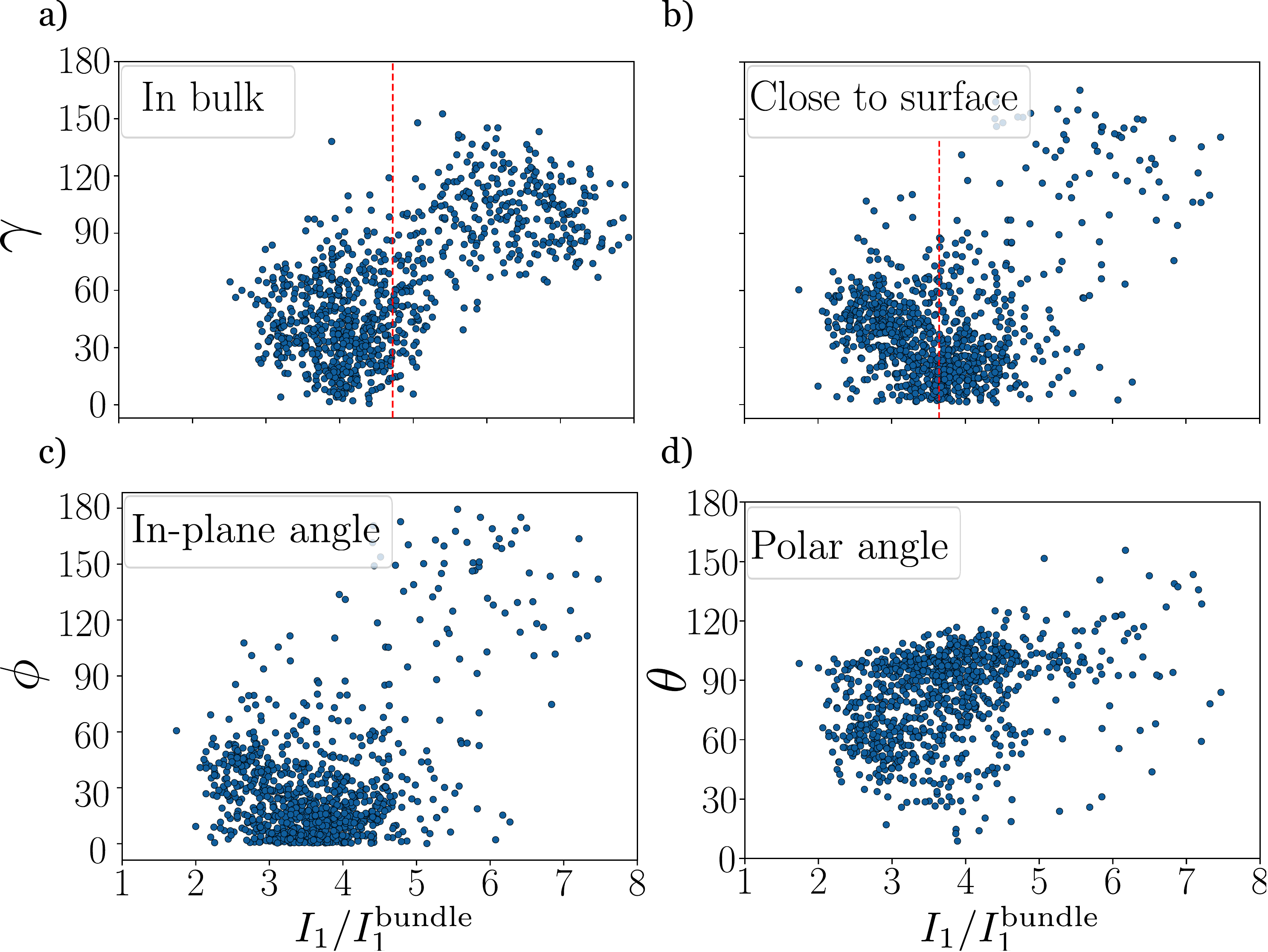}
  \caption{Several scatter plots correlating the reorientation of the bacterium due to tumbling with the time-averaged
  moment of inertia $I_1/I_1^\text{bundle}$: Tumble angle
  (a) in bulk and (b) near a surface, where the red dashed lines highlight the mean value of $I_1/I_1^\text{bundle}$. (c) In-plane angle and
   (d) polar angle.}
  \label{fgr:I_1_bulk_vs_surface}
\end{figure}

As in Section \ref{sec:analyze_tb_event_bulk}, we 
determine the smallest eigenvalue $I_1$ of the moment of inertia tensor of the four flagella and use it as a measure for
the dispersion of the flagellar bundle during tumbling. We correlate it with the tumble angle $\gamma$ in 
Figs.\ \ref{fgr:I_1_bulk_vs_surface}a) and b) that contain the scatter plots for the bulk fluid and close to a flat surface, respectively.
Supplemental video A8
shows a  tumble event with 
small dispersion $I_1/I^{\text{bundle}}_1 =2.5$ and a tumble angle $\gamma =35^{\circ}$,
while supplemental video A9
illustrates the case of
large dispersion 
with $I_1/I^{\text{bundle}}_1 =5.7$ and $\gamma =140^{\circ}$. 
Close to a surface, one still recognizes two regions in the scatter plot, but the tumble events with large dispersions and large 
tumble angles are less frequent. As a result,
the mean dispersion quantified by the mean value $\langle I_1 \rangle /I^{\text{bundle}}_1$
is significantly smaller near the surface.
In the bulk fluid, large
bundle dispersion predominantly 
is correlated with strong reorientation. 
This is still true near
a surface, 
but small tumble angles can also arise. 
In total, reorientations with smaller tumble angles dominate and are coupled to tighter bundles with lower dispersion. 
The reason
leading to reduced bundle dispersion 
close to the surface can have various origins. 
On the one hand, direct interactions between flagella and the surface certainly hinder the dispersion,
on the other hand, the enhanced drag due to the presence of a wall may also have an effect.

Figure \ref{fgr:I_1_bulk_vs_surface}c) shows the scatter plot relating in-plane reorientation angle and flagellar dispersion. It is comparable 
to the scatter plot of the tumble angle $\gamma$ in b), where small reorientation mainly correlates with small dispersion. In contrast, although 
less frequent at the surface, the large in-plane reorientation events are
mainly due to increased flagellar dispersion.
Nevertheless, they may 
help to explore the surface, which otherwise is limited
due to circular swimming.\cite{Lauga2006, Perez2019}

Finally, Fig.\ \ref{fgr:I_1_bulk_vs_surface}d) presents the scatter plot for the
polar angle. Contrary to expectation, 
large flagellar dispersion during tumbling is not correlated with
bacterial escape, since both low and high flagellar dispersions result
either in escape from ($\theta<90^\circ$)
or stay at ($\theta>90^\circ$)
the surface. The mechanisms to facilitate escape from the surface seem to be diverse. 
The \textit{E.\ coli} bacterium is a hydrodynamic pusher dipole, which is attracted and aligns parallel to the surface.\cite{Berke2008, Knut2011}
We argue that the dispersion of the flagellar bundle during tumbling weakens the pusher dipole and, therefore, the attraction towards the surface.
Furthermore, our videos reveal that the bacterium uses the direct contact of the flagella with the surface to push itself away from
the surface. However, a comprehensive analysis is necessary to elucidate the precise contribution of hydrodynamic and steric interactions.

\section{Discussion and conclusions}
\label{sec.conclusions}

In this article, we numerically investigated
an \textit{E.\ coli} bacterium
tumbling both in bulk and near a surface. More specifically, we examined the role of the surface during tumbling by systematically 
comparing our observations to those in
the bulk fluid.
We developed a fine-scale model that captures the mechanical properties of \textit{E.\ coli}, where flagellar flexibility and polymorphism are modeled using an extended version of Kirchhoff's rod theory. The modeled bacterium 
swims
in an MPCD fluid, which 
reproduces the flow field generated by the bacterium.
We simulated a specific 
setting
of an \textit{E.\ coli} with four flagella symmetrically attached to the rear of the cell body,
which allowed us to concentrate on the influence of tumble time, hook stiffness, and a bounding surface on the tumbling behavior.

First, we
investigated a long trajectory in 
the bulk fluid.
During 
tumbling we observe
a systematic drop in velocity followed by strong fluctuations. This velocity drop during tumbling is systematically observed in experiments.\cite{Berg1972, Turner2016, Lemelle2020} However, only ref. \citenum{Turner2016}
reports the observed velocity fluctuations, which they 
attribute to a
 motor reversal without a flagellum leaving the bundle. 
In our case, the fluctuations are caused by
 the erratic 
reorientation
of the cell body 
during tumbling.
We quantified reorientation after a single tumble event by calculating the 
the distribution of tumble angles and find
good agreement with 
experiments
from Berg and Brown (ref. \citenum{Berg1972}) in the case of a fixed tumble duration. 
Motivated by the experimental observations of ref. \citenum{Junot2022}, we determined the tumble angle distribution for a tumble 
duration
taken from a
Gamma distribution. Although the resulting tumble angle distribution does not
fully
reproduce experimental observations, we 
found that \textit{E.\ coli} needs to reversely rotate a flagellum for at least $10\tau_b$ to exhibit any 
noticeable
reorientation. 
Thus, such tumbling events
could be
overlooked in an experiment.

Another important feature of \textit{E.\ coli} is the flagellar hook, whose high flexibility enables the formation of the flagellar bundle. 
To study the influence of the hook's bending rigidity,
\cite{Nord2022, Zhang2023} we investigated bacterial tumbling with 
an increasingly stiffer hook. For a flexible hook, the bundle can become highly dispersed during tumbling,
which is also observed in experiments.\cite{Turner2000, Turner2016, Junot2022}. However,
increasing its bending rigidity systematically reduces flagellar dispersion, which we quantify by the smallest eigenvalue of the moment 
of inertia tensor. A stiffer hook 
hinders the rotation of both the reversely rotating flagellum and the cell body. We observe a more narrow
tumble angle distribution with a larger mean tumble angle compared to the reference case.
In particular, the reversely rotating flagellum protrudes more clearly from a tighter bundle, which agrees more with the traditional view of 
a tumble event. In refs. \citenum{Zhang2023} and \citenum{Wu-Zhang2025} it is argued that this enhances tumbling.
Still,
further experimental and theoretical investigations are needed to 
further elucidate the determining features of tumbling.

When immersed in a fluid bounded by two parallel no-slip walls, we 
observe significantly smaller tumble angles,
as also 
reported in several experiments.\cite{Molaei2014, Lemelle2020, Junot2022} 
A more detailed investigation reveals
a tendency for the bacterium to reorient parallel to the wall,
while the rate for swimming away from the surface is
$53\%$. The conditions leading to a successful escape are nontrivial.
One observation from our simulations is that  the bacterium seems to use direct contact between the flagella and the surface 
to push itself away from the wall.
Furthermore, during tumbling the pusher flow field, which attracts the \textit{E.\ coli} to the surface, is certainly weakened by the dispersed flagellar 
bundle. However, we did not find a clear correlation between the bacterial escape and bundle dispersion. Nevertheless, to decide whether 
hydrodynamic effects play a role for bacterial escape during tumbling, a detailed investigation of the flow field generated by \textit{E.\ coli}
is required, which is beyond what the current study can deliver.
We found a predominance of small in-plane reorientations near the surface. This observation could explain the apparent suppression of 
tumbling as
reported in ref. \citenum{Molaei2014}, 
since small in-plane reorientations
are harder to detect. 
Finally, a pronounced in-plane reorientation is correlated with a larger flagellar dispersion.

Our
model does not account for biophysical processes 
that govern
behavioral variability, as detailed in ref. \citenum{Junot2022}. 
Also, for our systematic study to explore the effect of a surface on tumbling, we modeled the bacterium with
four symmetrically attached flagella. 
However, the number and arrangement of the flagella varies largely. 
While 
this has
only a minor influence on swimming speed,\cite{Hu2015} 
the influence on tumbling needs further investigation.

In summary, we developed a detailed model of
an \textit{E. coli} bacterium
tumbling in bulk and near a surface. Our study provides new 
insights
into the role of flagellar dispersion and its influence on the tumble angle
distribution.
We found good agreement with experimental observations in bulk and suggest that the apparent suppression of tumbling near a surface
are due to small in-plane reorientation events that might have been overlooked
in experiments.
The versatility of our approach opens new avenues for further investigations. 
For example, the method of
MPCD facilitates the implementation of complex environments, such as 
different flow geometries like Poiseuille flow,  curved surfaces, and even model porous systems. Also,
the extended Kirchhoff-rod approach can be applied to study other peritrichous bacteria as well as different polymorphic flagellar
transformations.

\section*{Conflicts of interest}
There are no conflicts to declare.
\section*{Data availability}
The data supporting the findings of this study are available upon request from one of the authors.

\section*{Acknowledgements}
We thank Derek Cyrus Gomes, Arne Zantop and Zihan Tan for 
useful discussions on the development of the model
as well as Eric Clement and Katja Taute for helpful insights into the behavior of an \textit{E.\ coli} bacterium.
Financial support from the Deutsche Forschungsgemeinschaft (DFG) (Grant number 462445093) and TU Berlin
is gratefully acknowledged.


\balance


\bibliography{rsc} 
\bibliographystyle{rsc} 

\end{document}